\documentclass[twocolumn]{aa}
\usepackage{graphicx}
\begin{document}
   \title{$Chandra$ Detection of Hotspot and Knots of 3C~303}


   \author{Jun Kataoka
          \inst{1},
          Philip Edwards\inst{2},
          Markos Georganopoulos\inst{3},
          Fumio Takahara\inst{4}
          \and
          Stefan Wagner\inst{5}
   }

   \authorrunning{Kataoka et al.}

   \offprints{J. Kataoka}
   \institute{Tokyo Institute of Technology,
              Meguro-ku, Tokyo, Japan\\
              \email{kataoka@hp.phys.titech.ac.jp}
         \and
              Institute of Space and Astronautical Science, Sagamihara,
              Kanagawa, Japan
         \and
              Laboratory for High Energy Astrophysics, NASA/GSFC, Greenbelt,
              MD, USA
         \and
              Department of Earth and Space Science, Osaka University,
              Osaka, Japan
         \and
              Landessternwarte Heidelberg, K\"{o}nigstuhl,
              Heidelberg, Germany
             }

   \date{Received 17 October 2002 / Accepted 22 November 2002}

   \abstract{
We report the detection at X-rays of the radio/optical hotspot and
knots of 3C~303 from a short (15 ksec) $Chandra$ exposure in 2001
March. The X-ray morphology is similar to that of the radio/optical
emission with peaks in the X-ray emission found at 5.5$''$(knot~B), 9$''$
(knot~C) and 17$''$ (hotspot) from the core of 3C~303.
Despite the limited signal-to-noise ratio of the short $Chandra$
exposure, the X-ray photon spectrum was measured for the hotspot.
We construct the spectral energy distribution (SED) and find that
the X-ray flux is well below the extrapolation of the radio-to-optical
continuum, which we interpret as resulting from the production of X-rays
via inverse Compton scattering of both synchrotron photons (SSC) and
cosmic microwave background photons (EC/CMB).
The magnetic field strength, region size, and the maximum energy
of electrons are self-consistently determined for the hotspot
to be $B$ $\simeq$ 4.3 $\mu$G,  $R$ $\simeq$ 6.5$\times$10$^{21}$ cm,
and $\gamma_{\rm max}$ $\simeq$ 1.4$\times$10$^7$.
This implies a magnetic field strength  a factor of $\sim$ 30
below the equipartition value;  $B_{\rm eq}$ $\simeq$ 150 $\mu$G.
The origin of this large departure from equipartition is still uncertain,
but the discrepancy is reduced if the plasma in the hotspot is
moving with mildly relativistic speeds.
Our observation of 3C~303, as well as recent $Chandra$ detections of
large scale jets and hotspots in a number of radio galaxies,
confirm that particles are accelerated very efficiently in radio galaxies.

   }

   \maketitle
%

\section{Introduction}

It is generally believed that the vast bulk of cosmic-ray nuclei and
electrons up to 10$^{15}$~eV are accelerated in
the shock waves of supernova remnants (SNR).
There has been dramatic confirmation of this idea with the
detection of three SNRs as sources of $\sim$100~TeV particles,
both in the X-ray and TeV $\gamma$-ray bands;
SN~1006 (Koyama et al.\ 1995; Tanimori et al.\ 1998),
RXJ~1713.7$-$3946 (Koyama et al.\ 1997; Muraishi et al.\ 2000; Enomoto
et al.\ 2002) and
Cassiopeia~A (Hughes et al.\ 2000; Aharonian et al.\ 2001),
although it is still a mystery why only three of more than 200
known SNRs are TeV $\gamma$-ray emitters.
However, the cosmic ray spectrum extends well beyond 10$^{15}$~eV, and so
more powerful
particle accelerators must exist
in the universe. From the theory of Fermi acceleration, the maximum energy of
electrons/ions is proportional to the system size
$R$; $E_{\rm max}$ [TeV] $\sim$ 10$^{3}$ $R$ [pc] $B$ [$\mu$G] (Hillas 1984).
Thus the the location of strong shocks and more extended structures than
SNRs can potentially accelerate particles above 100~TeV.
The large scale jets
and lobes
in extragalactic sources are good candidates
for such acceleration sites (e.g., Hillas 1984; Gaisser 2000).

Jets are commonly observed in the radio galaxies, representing
outflows that are well collimated on scales of tens to hundreds
of kiloparsecs (kpc).
In spite of a long history of study,
however,
the processes by which the jets are formed and collimated are still far
from being understood. Detailed spatial imaging can potentially trace
the energy transport from the vicinity of the central black hole to the
outer jet region,
but observations with high spatial resolution ($\le 1''$) have
until recently only
been possible at radio and optical wavelengths.
Observations of jets at shorter wavelengths are extremely important because
they can probe the sites of higher energy particle acceleration.
Until the advent of $Chandra$, nonthermal X-ray emission had been
detected from only a few kpc-scale jets by  $Einstein$ and $ROSAT$
(e.g., Biretta, Stern \& Harris 1991; R\"{o}ser et al.\ 2000).
Marginal X-ray detections were reported also for the hotspots
in which the jets are believed to terminate
(Harris, Carilli \& Perley 1994; Harris, Leighly and Leahy 1998;).

The $Chandra$ X-ray Observatory has now resolved the X-ray spatial
structure along jets, from kpc to Mpc, of more than a dozen of radio
galaxies and quasars; e.g., PKS~0637$-$752 (Schwartz et al.\ 2000; Chartas et
al.\ 2000), M~87 (Marshall et al.\ 2002; Wilson \& Yang 2002),
3C~273 (Sambruna et al.\ 2001), and Cen~A (Kraft et al.\ 2002).
These observations have established that X-ray emission from
large scale jets is more common than had been expected. X-rays have also
been detected clearly from a number of lobes and hotspots
(e.g., Wilson, Young \& Shopbell 2000; 2001 for Pictor~A and Cygnus~A;
see Hardcastle et al.\ 2002 for a review).
X-ray observations of nonthermal emission from hotspots are
of particular interest to constrain the energetics and the spectrum
of the relativistic electrons. For example, the presence of
high energy electrons
with very short radiative life-times provides direct
evidence for re-acceleration in the hotspot.

3C~303 ($z$=0.141) is an unusual double radio source with a highly
asymmetrical structure. This source is classified as a broad-line
radio galaxy and had been thought to be a strong X-ray ``point source''.
In the $ROSAT$ HRI image a faint point-like component,
containing  25$\pm$6 counts, was found approximately coincident
with the western hotspot of 3C~303 (Hardcastle \& Worrall 1999).
However photon statistics and image resolution were not sufficient to
confirm that the X-ray detection was associated with the hotspot.
In particular, it was not possible to rule out a background ($z=1.57$)
QSO as the source of the X-rays (see below).
In this letter, we report on the $Chandra$ discovery of X-ray emission
from the hotspot and knots of 3C~303.
By combining the data from radio to X-rays, we show that radio
galaxies accelerate particles to energies of 10$-$100~TeV in the hotspot.
Throughout this paper, we adopt $H_0 = 75$~km\,s$^{-1}$\,Mpc$^{-1}$ and
$q_0 = 0.5$, so 1$''$ corresponds to 2.1~kpc at the redshift of
3C~303. 

\section{Observation and Data Analysis}

\subsection{Radio, Infrared, and Optical Observations}

The western hotspot of 3C~303 has been well studied at radio wavelengths.
A 1.5~GHz VLA image of 3C~303 is shown in Figure~1(a)
(Leahy \& Perley 1991). A prominent one-sided jet leads to a complex
radio hotspot (Lonsdale, Hartley-Davies \& Morison 1983) located at a
projected distance of 17$''$ (36~kpc) from the nucleus. The hotspot
is actually double (components ${\rm A_1}$ and ${\rm A_2}$
in the 408~MHz map of Lonsdale et al.\ 1983), with the
southern spot (${\rm A_2}$)
being the brighter in radio images. Also, three faint jet components
(knots~A, B, C in Figure~1(a)) are linearly spaced between the
nucleus and the hotspot.
The 1.5\,GHz image of Leahy and Perley (1991), made with the VLA in A and B
configurations, provided the first detection of the fainter hotspot on the
eastern side of the core. The fact that the eastern lobe is significantly
less polarized than the western lobe was ascribed to the Laing-Garrington
effect (Laing 1988; Garrington et al.\ 1988), confirming that the jet on
the western side is directed towards us, and that the counter-jet is on
the eastern side.


Recently, Giovannini et al.\ (2001) derived limits for the
angle to the line of sight, jet speed and Doppler factor of 24 radio
galaxies based on the jet--to--counter-jet ratio and core dominance.
For 3C~303, they derived a Doppler beaming factor in the range
$0.80 \le \delta \le 1.79$ by assuming a bulk Lorentz factor of
$\Gamma_{\rm BLK}$ = 5. They presented a single epoch 5\,GHz
VLBI image which shows  a number of parsec-scale jet components
(Figure~15 in Giovannini et al.\ 2001),
however, as no other VLBI images have been published, it is not
yet possible to determine the apparent jet speed directly.

   \begin{figure*}
   \centering
   \includegraphics[width=8.5cm]{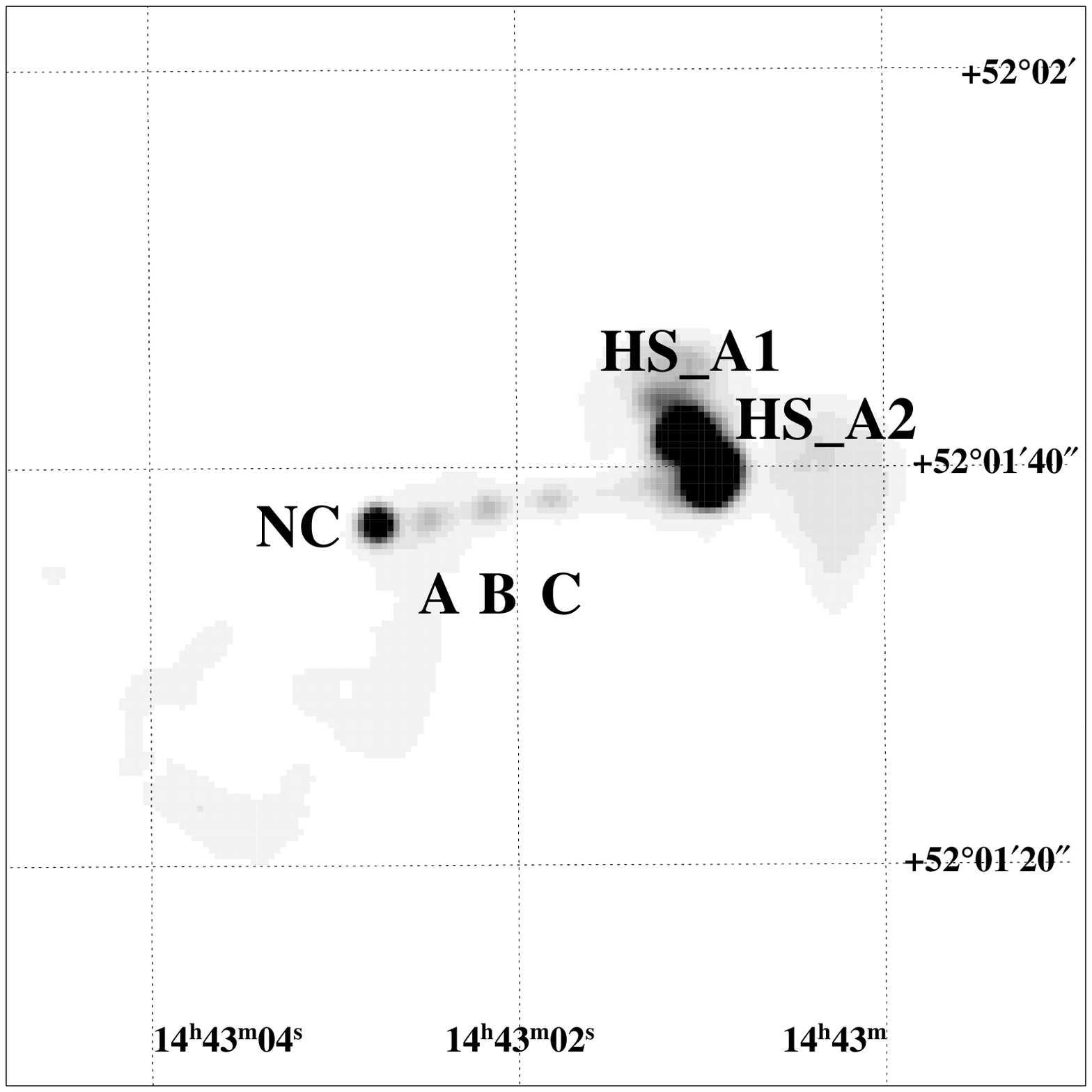}
   \includegraphics[width=8.5cm]{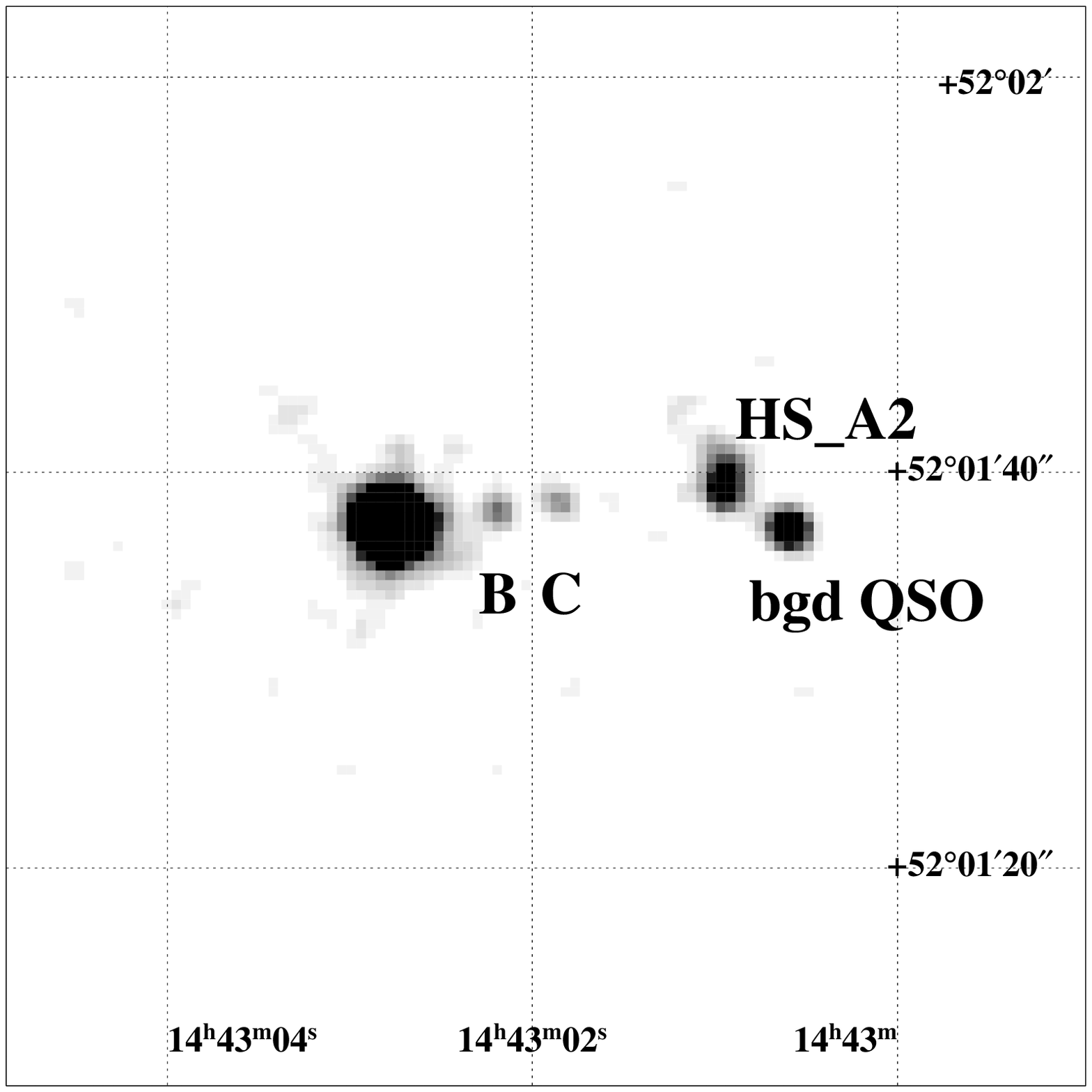}
      \caption{(a) Radio image of 3C~303. The grey scale is a
    1.5~GHz VLA image (Leahy \& Perley 1991; actually
    taken from Leahy, Bridle \& Strom 1998). The resolution is 1.2
    arcsec and black represents 20 mJy beam$^{-1}$.
    NC denotes the nucleus; A, B and C the jet knots;
    and HS$\_$A1 and HS$\_$A2 the hotspot components.
(b) X-ray image of
    3C~303 in the 0.4--8~keV band (ACIS-S onboard $Chandra$). The image is
    smoothed with a $\sigma$ = 0.5~arcsec Gaussian. 
    B and C denotes the jet knots,
    and HS$\_$A2 the hotspot component ${\rm A_2}$.
       }
         \label{image}
   \end{figure*}

An infrared image of 3C~303 is presented in Figure~1(g) of
Meisenheimer, Yates and R\"{o}ser (1997). They detected a clear signal
from the hotspot (${\rm A_2}$) and the jet which connects the nucleus
with the hotspot. The background QSO,
21$''$ from the nucleus and to south-west of the hotspot,
identified by Kronberg (1976) was also detected.

An optical counterpart of the multiple radio hotspot was found by
Leli\`{e}vre \& Wl\'{e}rick (1975) and Kronberg (1976).
Higher resolution images were obtained by
L\"{a}hteenm\"{a}ki \& Valtaoja (1999) using the 2.5\,m Nordic Optical
Telescope on La Palma in 1992 June.
They showed that the total extent of the hotspot, as measured
from the outermost 3$\sigma$ contours, is 2.3$''$ $\times$ 3.1$''$,
corresponding to a linear size of 4.9$-$6.6~kpc (without correction of
broadening effect of the optical PSF and the seeing of 0.6-0.8'').
The optical emission from the fainter northern hotspot
(${\rm A_1}$) was not detected,
indicating that radio-optical spectral indices are different in
the two components.  Considering the non-stellar structure of the image,
the positional agreement, and the optical polarization, they argue that
the optical candidate is a genuine counterpart to the radio hotspot
(${\rm A_2}$) of 3C~303.

\subsection{X-ray Observations}

$Chandra$ observed 3C~303 on 2001 March 23 in a guaranteed time
observation. The source was at the nominal aimpoint of the
Advanced CCD Imaging Spectrometer S3 chip (ACIS-S3), and the exposure
was continuous for 15.10~ksec. We have analyzed archival data on 3C~303
provided by HEASARC Browse
(http://heasarc.gsfc.nasa.gov/dp-perl/W3Browse/Browse.pl).
The raw level-1 data were reprocessed using the latest version
(CIAO~2.1) of the CXCDS software. We generated a clean data set by
selecting the standard grades (0, 2, 3, 4 and 6) and energy band
0.4$-$8~keV.

The X-ray image, produced by smoothing the raw $Chandra$ image with a
Gaussian of width 0.5$''$ in the energy range of 0.4$-$8 keV,
is shown in Figure~1(b). X-ray emission can be clearly seen from
the bright central core, the western jet knots, and the hotspot (${\rm A_2}$).
The total extent of the X-ray hotspot (${\rm A_2}$), measured from the
outermost of 3$\sigma$ contours in raw(un-smoothed) image, was $\sim 2.5''$.
This is consistent with the optical extent of the hotspot, though we do
not take into account the broadening of $Chandra$ PSF ($\sim 0.5''$
half-energy radius, but with significant ``wings'' at larger scales).
There is a suggestion of faint X-ray emission from the northern hotspot
${\rm A_1}$, but due to limited photon statistics
no conclusive statement can be made.
No emission was detected exceeding the background level
at the location of the weak eastern hotspot seen in the
Figure~15 of Leahy \& Perley (1991).

The object just to the south-west of the hotspot is the background QSO
at $z = 1.57$ mentioned previously (72 counts; 8.5 \,$\sigma$).
Knots~B and C are marginally detected at the 4.5\,$\sigma$ (20 counts)
and 4.4\,$\sigma$ (19 counts) levels,
respectively. The innermost knot (A in Figure 1(a)) cannot be
resolved due to the contamination from the bright nucleus.

   \begin{table*}
      \caption[]{Spectral fits to the 3C~303 knots, hotspot, and
the background QSO.}
         \label{tab1}
     $$
         \begin{array}{p{0.2\linewidth}clll}
            \hline
            \noalign{\smallskip}
            region   & {\rm image\hspace{2mm}radius}^{\mathrm{a}} & {\rm photon\hspace{2mm}index}^{\mathrm{b}}  & {\rm 2-10\hspace{2mm} keV\hspace{1mm}flux^{\mathrm{c}}} & {\rm red.\hspace{1mm}\chi^2\hspace{1mm} (dof)}\\
            \hline
            hotspot (A$_2$) & 3.0  & 1.4 \pm 0.2  & (3.7 \pm 0.5) \times 10^{-14}& 0.43(7)\\
            knot~B    & 1.5   & 2.0 (fixed)  & (3.5 \pm 1.2) \times 10^{-15}& 0.19(2) \\
            knot~C    & 1.5   & 2.0 (fixed)  & (3.5 \pm 1.2) \times 10^{-15}& 1.2 (1) \\
            QSO       & 1.5 & 2.0 \pm 0.3  & (1.6 \pm 0.3) \times 10^{-14}& 0.74 (6)\\
            \noalign{\smallskip}
            \hline
            nucleus$^{\mathrm{d}}$ & 180  & 1.65 \pm 0.05  & (2.6 \pm 0.2) \times 10^{-12}& 0.8 (29)\\
            nucleus$^{\mathrm{e}}$ & 180  & 1.60 \pm 0.08  & (2.6 \pm 0.3) \times 10^{-12}& 0.6 (13)\\
            \noalign{\smallskip}
            \hline

         \end{array}
     $$
\begin{list}{}{}
\item[$^{\mathrm{a}}$] Radius (in arcsec) of a circular region to extract X-ray counts.
\item[$^{\mathrm{b}}$] The best fit power-law photon index assuming a
                       Galactic $N_{\rm H}$ of 1.6$\times$10$^{20}$ cm$^{-2}$.
\item[$^{\mathrm{c}}$] In unit of erg cm$^{-2}$ s$^{-1}$.
\item[$^{\mathrm{d}}$] Analysis of archival $ASCA$ data (1995 May).
\item[$^{\mathrm{e}}$] Analysis of archival $ASCA$ data (1996 January).
\end{list}
   \end{table*}

The nucleus is a very bright point source, with the $>30$\%
photon pile-up preventing us from performing
any meaningful spectral or spatial analysis of this region.
As an alternative, we have analyzed archival $ASCA$ data to determine
the X-ray flux and the spectral shape of the nuclear emission.
$ASCA$ observed 3C~303 in 1995 May and 1996 January,
with a 20~ksec exposure each time. The X-ray fluxes were 2.6$\times$10$^{-12}$
erg cm$^{-2}$ s$^{-1}$ in both observations (Table~1).
No significant variation in the flux was detected between the two observations.
Note that, due to the image resolution of $\sim 3'$,
the jet, hotspot, and background QSO cannot
be separated from the nucleus for the $ASCA$ data,
however, as indicated in Table~1, the contribution of these
components to the nuclear flux is expected to be only $\sim 2\%$.
Also, the nuclear flux may have been changed between
$ASCA$ and $Chandra$ epochs. Despite these uncertainties, the
flux measured with $ASCA$ is consistent with that expected from the
piled-up fraction observed with $Chandra$.

In order to minimize the contamination from neighboring regions
for spectral fitting,
we extracted the X-ray emission from the hotspot
(${\rm A_2}$), knots~B and C, and the background QSO,
with circular regions of 3.0$''$, 1.5$''$, 1.5$''$, and 1.5$''$
radius, respectively.
Approximately 95$\%$ of counts are collected with a
circular region of 1.5$''$ radius for a point source.
Background subtractions were performed for each region, where the
the background counts were accumulated at the same off-nuclear distance
to the component regions.

The fitted spectra are summarized in Table~1.
Errors quoted in the table correspond to 1$\sigma$ uncertainties for
the parameter, unless otherwise stated.
In each case, we assumed a power-law function absorbed
by Galactic $N_{\rm H}$ only (fixed to 1.6$\times$10$^{20}$ cm$^{-2}$:
Stark et al.\ 1992). The hotspot and the background QSO were adequately
fitted with this model.
The X-ray photon index was $\Gamma_{\rm HS}$ = 1.4$\pm$0.2, and the
corresponding
2$-$10~keV flux was 3.7$\times$10$^{-14}$ erg
cm$^{-2}$ s$^{-1}$ for the hotspot.
Due to the limited photon statistics, we cannot determine the
spectra shape for knots~B and C. We thus fixed the spectral index to
$\Gamma_{\rm knot-B,C} = 2.0$ and estimated the flux.
The X-ray fluxes of the knots~B and C were 3.5$\times$10$^{-15}$ erg
cm$^{-2}$ s$^{-1}$ and 3.5$\times$10$^{-15}$ erg cm$^{-2}$ s$^{-1}$
in 2$-$10~keV band, respectively.

For the background QSO, $\Gamma_{\rm QSO}$ = 2.0$\pm$0.3, and the
corresponding 2$-$10~keV flux  was 1.6$\times$10$^{-14}$
erg cm$^{-2}$ s$^{-1}$. We remark again that although we have taken
great care in reducing the data, it is impossible to eliminate
the contamination of bright nucleus completely for knots~B and C.
Thus we mainly consider the hotspot (${\rm A_2}$)
in the following as it was detected at high significance
(92 counts; 9.6\,$\sigma$ level) and is well separated from the
nucleus (17$''$).

\section{Discussion and Conclusion}

Using a short (15~ksec) $Chandra$ observation, we have detected the X-ray
counterpart of the radio-optical hotspot (${\rm A_2}$) and knots~B and C
in the radio galaxy 3C~303.
Figure~2 shows the SEDs from radio to X-ray energies of the
western hotspot (${\rm A_2}$). The radio-to-optical data are taken from
Meisenheimer, Yates and R\"{o}ser (1997). This clearly indicates that the
X-ray flux obtained with $Chandra$ is well below the extrapolation
from the radio-to-optical continuum. While the radio-to-optical
continuum is generally believed to be synchrotron radiation,
the radiative mechanisms usually considered as the possible origin of
X-rays are either synchrotron (e.g., Pesce et
al.\ 2001; Marshall et al.\ 2002) or
inverse Compton scattering on various possible sources
of soft photons including the synchrotron photons
(SSC; e.g., Hardcastle et al.\ 2002) and the cosmic microwave background
(EC/CMB: e.g., Tavecchio et al.\ 2000; Sambruna et al.\ 2001).

   \begin{figure}
   \centering
   \includegraphics[width=8.5cm]{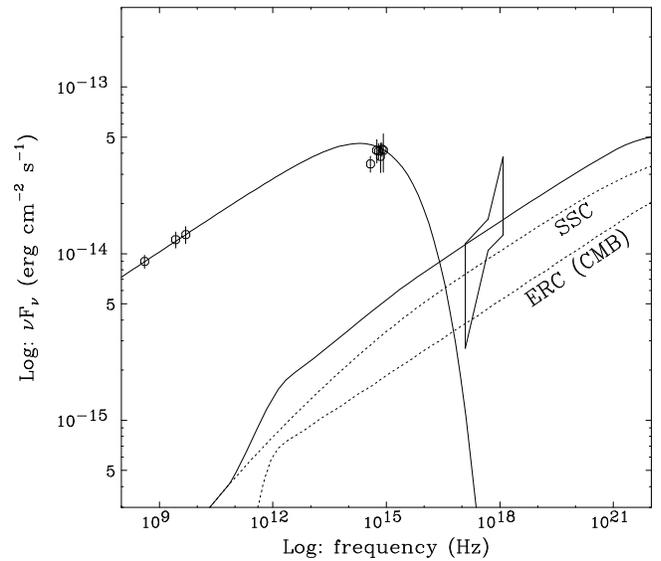}
      \caption{SED of hotspot(${\rm A_2}$) of 3C~303 (see Figure~1). The
    radio to optical data comes from Meisenheimer, Yates and R\"{o}ser 1997.
    The $bow$ $tie$ shows the X-ray data from $Chandra$ observation
    (this work; 90 $\%$ error in 0.5-5 keV).
    The solid line represents a fit assuming
    synchrotron emission for the radio to optical range and inverse Compton
    scattering of synchrotron (SSC) plus CMB photons (EC/CMB) for the
    X-ray emission, and yields $R$ = 6.5$\times$$10^{21}$ cm,
    $B$ = 4.3 $\mu$G and $\gamma_{\rm max}$ = 1.4$\times$10$^7$.
    Uncertainties in the physical parameters are discussed
    in the text and Figure~3.}
         \label{SED}
   \end{figure}

Relativistic boosting appears to be necessary to explain the X-ray
jet emission in the detected core-dominated radio-loud quasars.
Depending on the assumptions of jet Doppler factor,
different authors have reached different conclusions for
the physical processes to account for the SEDs
(compare, e.g., Schwartz et al.\ 2000 with Tavecchio et al.\ 2000).
However, the role of Doppler boosting in the case of hotspots
is still unclear.
Since hotspots are where the jet is thought to be terminated,
it is natural to consider that jet plasma is not beamed, or is
only mildly beamed. Furthermore, a statistical analysis
of samples of FR II radio galaxies implies non-relativistic velocities
in the hotspots (e.g., Arshakian \& Longair 2000). Taking the
results of Giovannini et al.\ (2001) into account, we start from
the assumption of a Doppler beaming factor of $\delta$ = 1.

Assuming a spherical geometry for the emission region,
the radius of the hotspot(${\rm A_2}$) is set to $R$ $\sim$
6.5$\times$10$^{21}$ cm ($\sim$ 1$''$; see $\S$ 2) from the radio, optical,
and X-ray images.
By fixing $R$ and $\delta$, other physical
parameters and the electron population are tightly constrained.
As described in the literature
(e.g., Kino, Takahara \& Kusunose 2002; Kataoka et al.\ 2002),
jet parameters are closely connected with the observed quantities.
The synchrotron cutoff frequency, $\nu_{\rm max}$,
is given by
\begin{equation}
\nu_{\rm max} \simeq 1.2\times 10^6 B \gamma_{\rm max}^2 (1+z)^{-1}\hspace{3mm}{\rm Hz},
\end{equation}
where $B$ is the magnetic field strength and
$\gamma_{\rm max}$$m_{\rm e}$$c^2$ is the
maximum electron energy.
Although the position of the synchrotron cut off is quite
uncertain, we infer that it must be between optical and X-ray energy
bands; $10^{14}$ Hz $\le$ $\nu_{\rm max}$ $\le$ $10^{16}$ Hz. This leads
to a condition in the $B$$-$$\gamma_{\rm max}$ plane,
\begin{equation}
9.5 \times 10^7 \le B \gamma_{\rm max}^2 \le 9.5 \times 10^9.
\end{equation}
The synchrotron luminosity is given by
\begin{equation}
L_{\rm sync} = 4\pi R^2 c U_{\rm sync}\hspace{3mm},
\end{equation}
where $R$ is the radius of the emission region, which we have
set to $R = 6.5 \times 10^{21}$\,cm.
We calculate the synchrotron luminosity integrated over all
frequencies using formula given by Band \& Grindlay (1985).
Assuming an electron population of the form
$N(\gamma)$=$N_0$$\gamma^{-s}$ exp($-$ $\gamma$/$\gamma_{\rm max}$), where
with the electron power law index set to $s$ = 2.7 from the radio
spectral index
$\alpha$ $\simeq$ 0.84 (Meisenheimer, Yates and R\"{o}ser 1997),
we obtain $L_{\rm sync}$ = 1.2$\times$10$^{43}$ erg s$^{-1}$.
Thus the synchrotron photon density is given as
$U_{\rm sync}$ = 1.7$\times$10$^{-12}$ erg cm$^{-3}$.
Comparing this with the cosmic microwave background(CMB) photon energy
density, $U_{\rm CMB}$ = 4.1$\times$10$^{-13}$(1 + $z$)$^4$  =
6.9 $\times$10$^{-13}$ erg cm$^{-3}$, the synchrotron photon density is
about a factor of two larger.
This indicates that the dominant source of seed photons which are
upscattered to the X-rays is synchrotron photons, but that CMB photons are
of comparable importance.

The ratio of the synchrotron luminosity to the SSC and EC/CMB
luminosities is given by
\begin{equation}
\frac{L_{\rm SSC} +L_{\rm EC/CMB}}{L_{\rm sync}} =  \frac{U_{\rm sync} + U_{\rm CMB}}{U_B},
\end{equation}
where $U_{B}$ is the magnetic field density ($U_B$ = $B^2$/8$\pi$).
Due to lack of $\gamma$-ray observations, we cannot measure Compton
luminosity ($L_{\rm SSC} +L_{\rm EC/CMB}$) accurately, however, it must be
larger than the ``X-ray luminosity'' which is defined in the
$Chandra$ bandpass ($L_{\rm X}$ $\equiv$ 4$\pi$$d_{\rm L}^2$$f_{\rm X}$ =
9.2$\times$10$^{41}$ erg s$^{-1}$, where $d_{\rm L}$ is the luminosity
distance to 3C~303 and $f_{\rm X}$ is the 0.4$-$8 keV flux).
This is because (i) the X-ray spectrum is still ``rising'' in the
$\nu$$F_{\nu}$ plane (see Figure 2) and (ii) the X-ray flux consider
only a small part (0.4$-$8 keV) of overall inverse Compton emission.
Considering this inequality, $L_{\rm SSC} +L_{\rm EC/CMB}$ $\ge$ $L_{\rm X}$,
the magnetic field is estimated as
\begin{equation}
B \le 28 \mu{\rm G}.
\end{equation}

 Combining equations (2) and (5), the ``allowed'' region of the
$B$$-$$\gamma_{\rm max}$ plane is shown as the hatched region in
Figure~3. The input parameters for our model are those at the
center of this region, and the resultant fit is given as solid line in
Figure~2. We assume an electron population of a form
$N(\gamma)$=$N_0$$\gamma^{-s}$ exp($-$ $\gamma$/$\gamma_{\rm max}$), where
$N_0$ = 0.9\,cm$^{-3}$, $\gamma_{\rm max}$=1.4$\times$10$^7$, and
$s=2.7$. We set the minimum Lorentz factor of electrons ($\gamma_{\rm
min}$) to 1, and the magnetic field to $B$ = 4.3\,$\mu$G, respectively.
Note that parameters we have chosen are valid even if the actual Compton
luminosity is an order of magnitude larger, in which case equation (5)
would be $B$ $\le$ 8.9 $\mu$G.

In Figure 3, we also show the region typical for  ``blazars''
for comparison (adapted from Ghisellini et al.\ 1998; Kubo et
al.\ 1998). Blazars are characterized by strong jet emission and the
radiation is thought to be emitted from a relativistic jet
directed close to our line of sight (e.g., Urry \& Padovani 1995).
More than 70 blazars are  strong GeV and/or TeV $\gamma$-ray emitters,
indicating that particles are accelerated very efficiently in sub-pc
scale jets up to $\gamma_{\rm max} \sim 10^6$ (e.g., Ghisellini et al.\
1998; Kubo et al.\ 1998).
Interestingly, our current analysis of 3C~303 implies that the
hotspots of radio galaxies may be $more$ powerful particle accelerators
than blazars, with particles being accelerated up to 10$-$100\,TeV.

Recent $Chandra$ and $ROSAT$ detections of large scale jets and
hotspots provide additional evidence for extremely relativistic particles in a
number of radio galaxies: M87 (Wilson \& Yang 2002:
$\gamma_{\rm max}$$\sim$10$^{7-8}$), 3C~390.3
(Harris, Leighly \& Leahy 1998: $\gamma_{\rm max}$$\sim$7.5$\times$10$^7$),
and 3C~371 (Pesce et al.\ 2001: $\gamma_{\rm
max}$$\sim$4$\times$10$^7$). Similar characteristics, such as steep spectral
indices and high-frequency cutoffs, have also been reported
for a number of hotspots in other sources
(3C~123, Pictor A, 3C~295, 3C~351; see Harris and Krawczynski 2002).
Although the sample size is still small, it appears that large scale
jets/hotspots in radio galaxies are an important acceleration
site in the universe.

If the electrons are actually accelerated
to such high energies, the
electrons emitting via synchrotron in the optical band and via inverse
Compton at $\gamma$-ray energies have relatively short lifetimes.
The cooling time of electrons is
\begin{equation}
t_{\rm cool}(\gamma) = \frac{3 m_{\rm e}c}{4 (U_{\rm B}+U_{\rm sync}+U_{\rm CMB})\sigma_T \gamma} .
\end{equation}
Assuming the parameters used to model the SED,
$\gamma_{\rm max}$ = 1.4$\times$10$^7$ and
$B$ = 4.3$\mu$G,
we obtain $t_{\rm cool}$($\gamma_{\rm max}$) = 7.0$\times$10$^{11}$\,sec.
In other words, the highest energy electrons can travel
$d \sim c t_{\rm cool} = 6.8$\,kpc before losing their energy.
This is consistent with the size of the hotspot, but
smaller than the projected distance from the nucleus (36~kpc),
implying the electrons can originate
from a single acceleration site in the hotspot.
However considering the uncertainties in the $B$$-$$\gamma_{\rm max}$ plane,
we cannot establish with this data alone that electrons entering in the
hotspot actually need to be re-accelerated.

   \begin{figure}
   \centering
   \includegraphics[width=8.5cm]{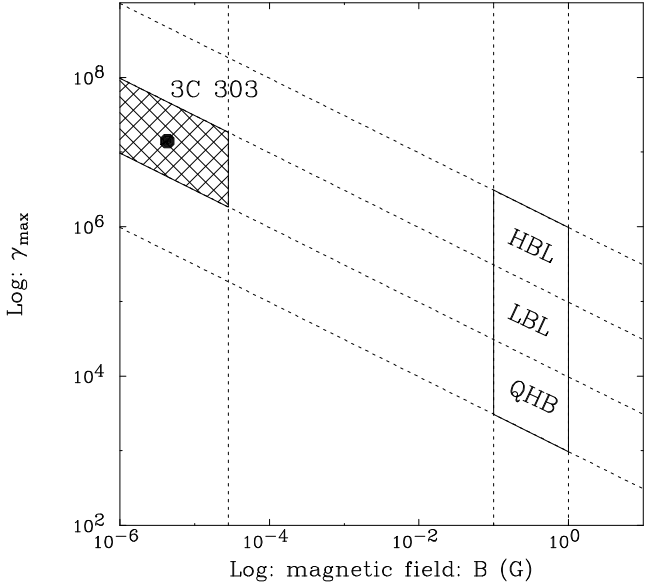}
      \caption{The parameter space ($B$, $\gamma_{\rm max}$) allowed by
the one-zone synchrotron plus SSC+EC/CMB model for the 3C~303 hotspot.
Filled circle represents parameters which fit the SED (see Figure 2).
The allowed parameter space for sub-pc scale jets in blazars
are shown for comparison. HBL; High Frequency Peaked BL Lac,
LBL; Low Frequency Peaked BL Lac, and QHB; QSO hosted blazars
(adapted from Kubo et al.\ 1998 and Ghisellini et al.\ 1998). }
         \label{BGreg}
   \end{figure}

One difficulty of understanding the overall photon spectrum with the above
simple picture is that the relativistic electron energy density
significantly exceeds the magnetic field density
($U_{\rm e}$ $\gg$ $U_B$).
For the parameters used to fit the SED (Figure~2), we expect
$U_{\rm e}$/$U_B$ $\simeq$ 1.4$\times$10$^6$$\gamma_{\rm min}^{-0.7}$,
where $\gamma_{\rm min}$ is the minimum Lorentz factor of electrons.
An equipartition magnetic field  $B_{\rm eq}$ $\sim$ 150 $\mu$ G
($B_{\rm eq}$/$B$ $\ge$ 30), underproduces the X-ray flux.
Such  large departures from equipartition have been suggested for other
X-ray hotspot sources, namely Pictor~A ($B_{\rm eq}$/$B$ $\simeq$
14; Wilson, Young \& Shopbell 2001) and 3C 351 ($B_{\rm eq}$/$B$
$\simeq$ 12 for the northern hotspot, and $<$25 for the southern hotspot;
Hardcastle et al.\ 2002).

The process responsible for the anomalously bright, flat X-ray
hotspots in these sources is still an open question.
Hardcastle et al.\ (2002) have suggested a number of possibilities,
some of which may also be applicable for 3C~303 hotspots.
One idea is that the hotspot is mildly beamed rather than
stationary. Beaming causes the observed flux densities to change by
a factor $\delta$$^{3+\alpha}$ (Dermer \& Schlickeiser 1993),
where $\delta$ is the  Doppler beaming factor.
For an electron population with energy index $s$, the predicted
SSC flux scales as $\delta$$^{-12/(s+5)}$ $\simeq$ $\delta$$^{-1.6}$,
and so Doppler dimming ($\delta$ $<$1) increases the predicted
equipartition flux. To place 3C~303 at
equipartition with the beamed SSC emission,
we thus require $\delta$$\sim$~0.01.
This strongly conflicts with the range
derived from radio observations,
$0.80 \le \delta \le 1.79$ (Giovannini et al.\ 2001),
although these limits were based in part on the properties of
the parsec-scale jet, and thus it may be
possible that the downstream hotspot has quite a
different Doppler factor.

For the case of 3C~303 hotspot, however, considering the SSC
emission alone  is not  sufficient because the synchrotron photon energy
density only marginally dominates over the CMB photon energy density
(by about a factor of two; see above).
For a source in equipartition the normalization factor of
the electron number density scales as
$N_0 \propto L_{\rm s} B_{\rm eq }^{- (1+\alpha)}$
(Longair 1994), where $ L_{\rm s}$ is the synchrotron comoving luminosity at a
fixed frequency for a source of a given comoving volume.
The equipartition magnetic field is
$  B_{\rm eq } \propto L_{\rm s}^{2/7}$.
Combining these two equations, we obtain
$N_0\propto L_{\rm s}^{(5-2\alpha)/7}$.
The observed synchrotron luminosity at the same fixed frequency is
$L_{\rm obs}=L_{\rm s} \delta^{3+\alpha}$.
Using this, we obtain
$N_0 \propto \delta^{-(3+\alpha)(5-2\alpha)/7}$.
Therefore, for a given observed synchrotron spectrum,
as  the beaming increases, the  electron number density in equipartition
decreases.
The external Compton emission is
$L_{\rm EC} \propto N_0 \delta^{4+2\alpha}$
(Dermer 1995; Georganopoulos, Kirk \& Mastichiadis 2001).
Substituting $N_0$ we obtain
$ L_{\rm EC} \propto  \delta^{(2\alpha^2+15\alpha+13)/7}$.
Therefore, although the electron number density decreases as $\delta$
increases, the external Compton emission increases. Setting $\alpha=0.84$,
for a hotspot Lorentz factor $\Gamma_{\rm BLK}=2$
at an angle $\theta=30^\circ$ the external
Compton emission is boosted by a factor of $\sim 15$.
Such hotspot Lorentz factors are routinely observed in
hydrodynamic simulations (e.g., Aloy et al.\ 1999).

In conclusion, we suggest  that  for typical orientation angles
of $\sim 30^\circ$, the hotspot of 3C~303 would be closer
to equipartition if it were mildly beamed, but it is not clear if beaming
alone can account for  the large departure from equipartition.
Many theoretical models are now being developed, involving the hydrodynamic
simulations and MHD calculations, to understand the physical conditions
in hotspots
(Aloy et al.\ 1999; Tregillis, Jones \& Ryu 2001; Saxton et al.\ 2002).
These numerical simulations lead to us to expect that hotspots are
transient structures, and may well have different magnetic field and
electrons acceleration behaviour at different times. If this is the
case, it might be possible that the hotspots are being in a temporary
transition state, far from equipartition.

We finally comment on the X-ray emissions  from knots~B and C, which
may be compared to the hotspot(A$_2$) emission.
Our preliminary analysis shows that the radio flux of knots~B and C at
1.5\,GHz is $\sim$ 6$\times$10$^{-16}$ erg cm$^{-2}$ s$^{-1}$, which
is about five times fainter than the corresponding X-ray fluxes
(Table~1). The flux ratio between the radio and X-ray bands is almost
equal to that of the hotspots A$_{\rm 2}$ ($f_{\rm X}$/$f_{\rm radio}$
$\simeq$ 4).
This may imply that the SED of knots~B and C could be similar to
that of the hotspot, although the knots are
a~priori more likely to be moving at
relativistic speeds than hotspot A$_{\rm 2}$.
Unfortunately, optical detections of knots~B and C have not yet been
reported and so we cannot confirm this suggestion at present.
Deeper $Chandra$ and $HST$
observations of radio galaxies  will allow us to
understand the kinematics of the jet knots
hotspot and further study their likely role
as one of the most efficient particle
accelerators in the universe.

\begin{acknowledgements}
We deeply appreciate the insightful comments and suggestions of
the referee, M.J.\ Hardcastle, that have improved this paper
considerably.  J.K.\ acknowledges a Fellowship of the Japan Society
for Promotion of Science for Japanese Young Scientists.
\end{acknowledgements}

\end{document}